*Note to the readers: the manuscript was submitted to a journal for publication on June 3, 2016.*

# Near-infrared luminescent phosphors enabled by topotactic reduction of bismuth-activated red-emitting crystals

B.-M. Liu,[a] Z.-J. Yong,[a] Y. Zhou,[a] D.-D. Zhou,[a] L.-R. Zheng,[b] L.-N. Li,[c] H.-M. Yu,[d] and H.-T. Sun*[a]

[a.] College of Chemistry, Chemical Engineering and Materials Science, State and Local Joint Engineering Laboratory for Novel Functional Polymeric Materials, Soochow University, Suzhou 215123, China.  Email: timothyhsun@gmail.com

[b.] Synchrotron Radiation Facility, Institute of High Energy Physics, Chinese Academy of Sciences, Beijing 100049, China

[c.] Synchrotron Radiation Facility, Shanghai Institute of Applied Physics, Chinese Academy of Sciences, Shanghai 201204, China

[d.] Shanghai Institute of Ceramics, Chinese Academy of Sciences, Shanghai 200050, China

## Abstract

Bismuth-doped luminescent materials have gained significant attention in the past years owing to their huge potential for the applications in telecommunications, biomedicine, and displays. However, the controlled synthesis of these materials, in particular for those luminescing in the near-infrared (NIR), remains a challenging subject of continuous research effort. Herein, we show that the low-temperature topotactic reduction by using Al metal powders as oxygen getters can be adopted as a powerful technique for the conversion of bismuth-doped red-emitting systems into NIR-emitting cousins as a result of the creation of unique crystalline networks. Thorough experimental characterization indicates that the framework oxygen of hosts can be topotactically extracted, thus producing unique metal-oxygen-metal networks in the reduced phases while preserving the crystalline structure of the precursor. Furthermore, X-ray absorption spectroscopy reveals that Bi atoms substitute for both $Ba^{2+}$ and $P^{5+}/B^{3+}$ in $BaBPO_5$ crystals, and subsequent topotactic treatment preferentially changes the local environment of Bi at $P^{5+}/B^{3+}$ sites, which results in the occurrence of NIR emission owing to the creation of NIR-luminescent, defective Bi-O polyhedra in which Bi bears lower oxidation states with respect to that in the precursor. The site-specific topotactic reduction reaction reported here helps us create peculiar NIR-luminescent Bi-O units, and simultaneously does not seriously affect the red photoluminescence from $Bi^{2+}$ situated at the $Ba^{2+}$ sites. Given the long-lived, ultrawide NIR emission covering the second biological windows, the phosphors developed here hold great promise for *in vivo* luminescence and lifetime bioimaging. We anticipate that this low-temperature topotactic reduction strategy can be applied to the development of more novel Bi-doped luminescent materials in various forms that can find a broad range of functional applications.

## Introduction

Over the past decades, photonic materials activated by bismuth have been of intense interest owing to their tremendous potential for the applications in the fields of telecommunications, biomedicine, and displays.[1-26] Up to now, it is well recognized that $Bi^{3+}$ and $Bi^{2+}$-doped materials could demonstrate photoluminescence (PL) in the UV and visible spectral range.[4] Concurrently, some systems containing Bi have been found to exhibit peculiar infrared PL that drastically differs from $Bi^{3+}$ and $Bi^{2+}$-doped cousins, and a diverse array of such materials in various forms have been developed, including bulk glasses and crystals,[5,8] glass fibers and films,[6b,7] zeolites,[9,11] molecular crystals,[13-16] ionic liquids,[12] and nanoparticles (ref.10). Interestingly, fiber lasers, optical amplifiers and bioimaging have been successfully demonstrated by using corresponding systems because of the tunable and high-efficiency infrared emission characteristics.[6,10] However, the techniques available for the synthesis of this class of materials are rather limited and specific.[4] Although near-infrared (NIR) PL has been observed in Bi-activated glasses and crystalline compounds, unfortunately, the preparation of these materials commonly relies on the high-temperature strategy, which renders it impossible to monitor the valence evolution from typical $Bi^{3+}$ ions in the precursors to NIR-active species in the final products; additionally, their PL properties are very sensitive to the preparation processes and Bi tends to form metallic nanoparticles in glasses prepared by a melt-quenching method, leading to difficulty in the rational design of these materials owing to the poor establishment of structure-property relationship.[4] Although wet-chemistry routes can be used for the synthesis of Bi-bearing molecular crystals with uncontested PL origins,[13-16] unfortunately, they are commonly material specific.

The diverse and complex NIR PL behaviors exhibited by Bi-doped compounds can be commonly attributed to the presence of *peculiar Bi-anion units* that are optically active; in most cases, the anion is oxygen. Based on this design discipline, an effective strategy for the discovery of new materials containing Bi which exhibit novel optical behavior is the preparation of phases containing *unusual Bi-anion networks*.[21,27] Recently, there has been a growing interest in manipulating and tuning the structures and physical behavior of compounds by modifying their anion lattices.[28-46] One of the simplest ways to accomplish this is by lowering the oxygen stoichiometry of materials, resulting in the formation of anion-deficient phases.[35] The decreased oxygen stoichiometry in these phases leads to a reduction in the average oxidation states of metals present. Most importantly, anion-deficient phases contain vacant anion sites which lower metal coordination numbers and disrupt metal-anion-metal coupling networks. These synergistic effects can radically change the magnetic coupling, electronic transport, as well as photophysical behaviors of a targeted material system.[35] One of the most powerful routes to anion-deficient materials is low-temperature reduction. Because anions were extracted at a sufficiently low temperature, there is insufficient thermal energy to allow the cations to response to changes in anion lattices, thus conserving the structure and cation order or disorder of the precursors used.[35] This strategy could produce numerous new metastable phases bearing peculiar magnetic and optical features, owing to the creation of *unique metal-anion networks* that are hard to be accessible by using conventional high-temperature reactions.[28-46]

Recently, Sun and coworkers found that such a defect-engineering approach can be adopted as a universal method to topotactically convert $Bi^{3+}$-doped phosphors into NIR-emitting ones.[21] By utilizing *calcium hydride (CaH$_2$)* as a low-temperature reducing agent, it is feasible to obtain a broad spectrum of Bi-doped metastable phases that contain randomly-distributed oxygen vacancies, thus resulting in the occurrence of under-coordinated Bi-O polyhedra that give rise to NIR PL. One of the most important attributes of this technique lies in the successful connection of traditional $Bi^{3+}$-doped systems with NIR-emitting ones in a rational and controllable manner, greatly deepening the understanding of Bi-related photophysical behavior and opening up new avenues to develop novel luminescent systems that may be promising for various functional applications. As is well recognized, Bi ions bear quite different coordination environments in Bi-doped red-emitting systems with respect to $Bi^{3+}$-doped cousins, demonstrating red PL with featured excitation bands that is commonly attributed to $Bi^{2+}$ ion.[24-26] Inspired by the recent advance in structure-directed PL tuning by employing $Bi^{3+}$-doped phosphors as precursors, one interesting and practical question arises: is it possible to topotactically convert $Bi^{2+}$-doped phosphors into infrared-luminescent materials at low temperatures? If this could succeed, we thus can bridge the gap between $Bi^{2+}$-doped materials and NIR-luminescent counterparts, which will not only enrich the strategies for the synthesis of new generation of Bi-doped luminescent materials, but also provide unsurpassed insight into PL mechanisms of Bi-doped systems, in particular for those luminescing in the infrared.

In this contribution, we demonstrate that the low-temperature topotactic reduction strategy can be applied to the conversion of Bi-doped red-emitting systems into NIR-luminescent materials. We select a well-studied Bi-doped phosphor (abbreviated as Bi:BaBPO$_5$) as a precursor, the oxygen of which is extracted by using Al metal powders instead of CaH$_2$ as oxygen getters. The structural and photophysical properties of Bi:BaBPO$_5$ and the resulting reduced products were thoroughly examined by using steady-state and time-resolved PL, X-ray diffraction (XRD), thermogravimetric (TG) analysis, Raman scattering, and X-ray absorption fine structure (XAFS) spectroscopy. Experimental results reveal that the framework oxygen of the $BaBPO_5$ phase can be topotactically extracted, thus producing unique metal-oxygen-metal networks in the reduced phases while conserving the crystalline structure of the precursor. Furthermore, X-ray absorption spectroscopy verifies that Bi atoms substitute for $Ba^{2+}$ and $P^{5+}/B^{3+}$ in $BaBPO_5$ crystals, and subsequent topotactic treatment preferentially changes the local environment of Bi at $P^{5+}/B^{3+}$ sites, resulting in the occurrence of NIR emission owing to the creation of defective Bi-O units. The site-specific topotactic reduction reaction strategy helps us create peculiar NIR-luminescent Bi-O units, and simultaneously does not seriously affect the red PL from $Bi^{2+}$ occupied the $Ba^{2+}$ sites.

## Experimental section

**Synthesis of the precursors:** Bi-doped polycrystalline $BaBPO_5$ (1mol% Bi nominal composition) was synthesized using a solid state reaction method. Stoichiometric amounts of $BaCO_3$ (Aladdin, 99.99%), $H_3BO_3$ (3mol% excess, Aladdin, 99.99%), $NH_4H_2PO_4$ (3mol% excess, Aladdin, >99.99%) and $Bi_2O_3$ (Aladdin, 99.99%) were thoroughly mixed and ground. The mixture was first sintered at 500 °C for 5 h in air. The resulting powders were then reground, and finally sintered at 950 °C in air for 8 h, producing white powders.

**Reduction of the precursors:** Approximately 0.6 g of the precursor ($Bi:BaBPO_5$) was placed in a small semi-closed glass tube and then sited in a larger and longer glass tube with one mole equivalent of Al powder (Aladdin, 99.95%) acting as an oxygen getter in the bottom. All operations are carried out in a nitrogen-filled glovebox ($O_2$ and $H_2O$ < 0.1 ppm) in order to prevent the oxidation of Al powder. The larger tube was then flame sealed under vacuum. The tubes were then heated at temperatures ranging from 450 to 570 °C for different duration. The samples were denoted BBPO@T-t, where T and t represent the treatment temperature in degrees centigrade and duration in hours, respectively.

**Characterization:** Powder XRD measurements were carried out on a desktop diffractometer (D2 PHASER, Bruker, λ = 1.54056 Å) radiation operated at 30 kV and 10 mA. Photoluminescence spectra were recorded with an FLS 980 spectrofluorometer (Edinburgh Instruments Ltd.). The red PL spectra were analyzed using a monochromator (Horiba, iHR550) and detected using an electrically cooled PMT (Hamamatsu, R928) under the excitation of a 407 nm laser diode. Time-resolved PL measurements were performed by detecting the modulated luminescence signal with a PMT (Hamamatsu, H10330-75), and then analyzing the signal with a photon-counting multichannel scaler. The excitation source for the lifetime measurements was 476 nm light from an optical parametric oscillator pumped by the third harmonic of a Nd:YAG laser. Absorbance spectra of the samples were taken by a UV–vis-NIR spectrophotometer (Cary 5000, Agilent) equipped with an integrating sphere. TG measurements were performed by heating powder samples under an oxygen atmosphere with a flow rate of 40 ml min$^{-1}$ and measuring the mass change as a function of temperature using a Netzsch STA 449C thermal analyzer. Approximately 80 mg of the sample was loaded into an alumina crucible and heated at 10 K min$^{-1}$ from 20 to 800 °C. Raman scattering measurements were carried out by means of a Horiba XploRA micro-Raman spectrometer using the 532 nm laser line (5mW). The excitation line was focused on the samples through the 50 × long-working-length objective. The backscattered radiation was collected by the same microscope optics and dispersed by a monochromator equipped with a 1800 line/mm holographic grating. The sizes of the slit and pinhole were 100 and 300 μm, respectively. The dispersed radiation was detected by means of a Peltier-cooled charged-coupled device (CCD) sensor. The X-ray absorption spectra of the Bi $L_{III}$ edge for all samples were obtained on the 1W1B beam line of the Beijing Synchrotron Radiation Facility with stored electron energy of 2.5 GeV and average ring currents of 200 mA. A fixed-exit Si (1 1 1) double crystal monochromator was used. Note that the X-ray absorption spectra of these samples were also taken at the XAFS station (beam line BL14W1) of the Shanghai Synchrotron Irradiation Facility with stored electron energy of 3.5 GeV and ring currents of 200 mA. The results taken at both facilities are almost identical. Before XAFS measurement, the powder samples were dispersed on a single-faced adhesive tape or pressed into thin discs to adjust the thickness of the samples. $β-Bi_2O_3$ and Bi powder were used as reference samples. Before taking measurement, Bi metal was thermally treated at 150 °C for 1h in a reducing atmosphere ($H_2/N_2$: 5%) to remove the oxide layer on the surface. Data were collected in the fluorescence mode for the $BBPO_5$:Bi sample and in the transmission mode for the $β-Bi_2O_3$ and Bi powder.

## Results and discussion

The topotactic reduction of Bi:$BaBPO_5$ was carried out by using Al powders as oxygen getters. The XRD pattern of the as-synthesized Bi:$BaBPO_5$ can be readily indexed with a hexagonal unit cell (JCPDF no.

19-0096) (**Fig. 1**). The compound is built up of infinite anionic units of $[BO_4]$ and $[PO_4]$ tetrahedra joined by common vertexes, forming single chains of $[BO_4]$ tetrahedra running parallel to [001] that are linked to terminal $[PO_4]$ tetrahedra to form spiral chains. $Ba^{2+}$ ions are ten-fold coordinated with oxygen atoms. After thermal treatment, the resulting products obtained under different preparation conditions show similar diffraction patterns to the precursor, indicating that the crystalline structures maintain well.

**Fig. 2** displays the PL spectra of the as-synthesized and thermally-treated Bi:BaBPO$_5$ phosphors. The as-synthesized powders demonstrate a red emission band centered at 642 nm (Fig. 2a), which can be well ascribed to the electronic transition, $^2P_{3/2}(1) \rightarrow {}^2P_{1/2}$, of $Bi^{2+}$ ions. Obviously, thermal treatment of Bi:BaBPO$_5$ results in weaker red PL, accompanied by the appearance of a new NIR PL band peaked at ca. 1164 nm with typical excitation bands at 478 and 710 nm (Fig. 2b and Fig. S1). We note that the red PL intensity of the BBPO@450-x sample is irregular, and the BBPO@510-24 sample possesses the weakest emission. We will discuss this issue below. It is also find that the BBPO@450-3 sample demonstrates the strongest NIR PL, and the BBPO@570-24 almost does not show any NIR emission. All these suggest that the mild treatment condition is of paramount importance for achieving the strongest NIR emission. Furthermore, the undoped BaBPO$_5$ does not show this class of NIR emission. All PL results imply that the NIR emission originates from Bi-related active centers created by the thermal treatment of Bi:BaBPO$_5$. We next took the excitation-emission matrix (EEM) spectrum of the treated sample to know more details of the PL features. Notably, the EEM graph demonstrates two peaks corresponding to the emission/excitation wavelengths (unit: nm) of 1160/476 and 1120/700 (Fig. 3a), strongly indicating that the NIR emission stems from more than one class of optically active centers. Time-resolved PL measurement indicates that all treated products show multi-exponential decays (Fig. 3b). The effective lifetimes at 1164 nm are in the range of 81-116 μs (Table S1). In addition, it is found that the thermal treatment results in the occurrence of a broad absorption tail ranging from 320 to 1000 nm (Fig. 4); with the increase of treatment duration, the absorption becomes more intensive.

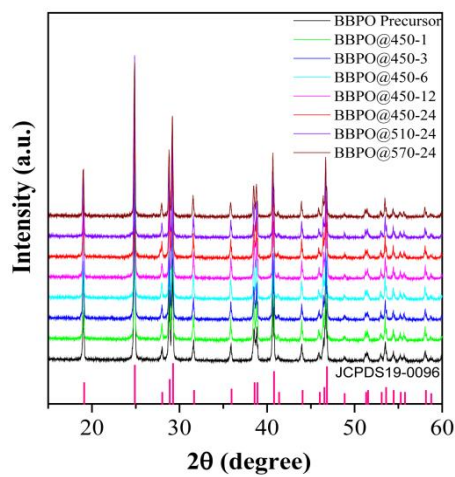

**Fig. 1** XRD patterns of the precursor and reduced products.

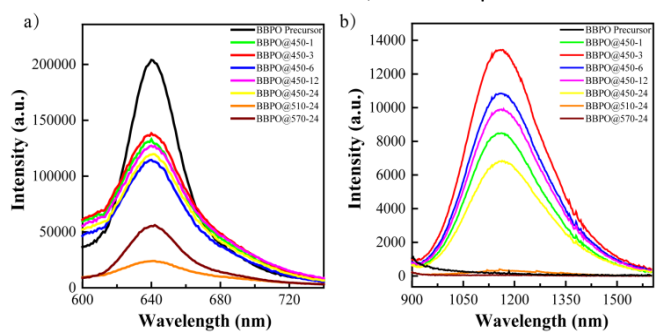

**Fig. 2** PL spectra of the precursor and reduced samples. (a) Visible emission spectra excited at 407nm. (b) NIR emission spectra excited at 476nm.

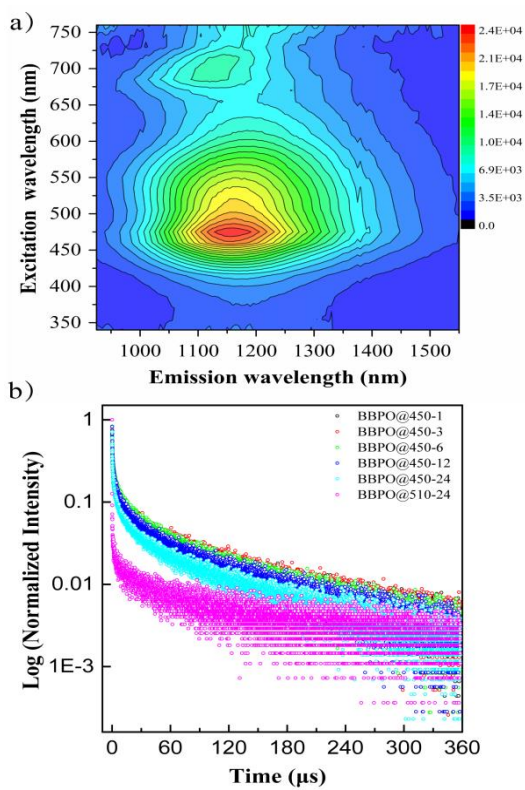

**Fig. 3** (a) Excitation–emission graph of the BBPO@450-3 sample. (b) Decay curves for the reduced samples. Measurements were performed by monitoring the decay of the emission at 1164 nm upon 476 nm excitation.

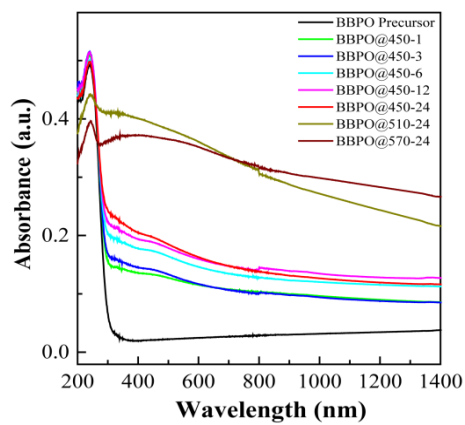

**Fig. 4** Absorption spectra of the precursor and reduced samples.

It is envisaged that the Bi:BaBPO$_5$ could release oxygen upon thermal treatment in vacuum, which reacts with Al getters to form the thermodynamically stable Al$_2$O$_3$. The extraction of oxygen is a temperature- and time-dependent process that allows the facile control of the amount of oxygen vacancies created. To verify this, we next carried out the TG analysis to compare compositions before and after thermal treatment. The typical TG data were collected by heating the treated powders in flowing oxygen for the evaluation of their oxygen stoichiometry (Fig. 5). It is obvious that the BBPO@450-6 sample displays a weight loss of 0.26%, which can be explained by the loss of absorbed water molecules and remnant B-related byproducts since 3% excess of H$_3$BO$_3$ was used for the synthesis of Bi:BaBPO$_5$. Notably, a slight weight increase is observed for the BBPO@570-24 sample after 500 ℃. Assuming the BBPO@450-6 does not have any oxygen vacancy and the byproducts are same in two samples, we can roughly deduce the composition of the BBPO@570-24 as Bi:BaBPO$_{4.98}$. This suggests that only a minority of oxygen atoms in the host were extracted by the thermal treatment. It is noted that the BBPO@570-24 sample does not show any NIR PL, along with the weaker red emission, giving an indication that too many oxygen vacancies can destroy the NIR active center while cannot strongly influence the Bi$^{2+}$ emitters (Fig. 2). We thus conclude that the NIR active centers are not mainly converted from Bi$^{2+}$ emitters, but from other classes of Bi in the matrix.

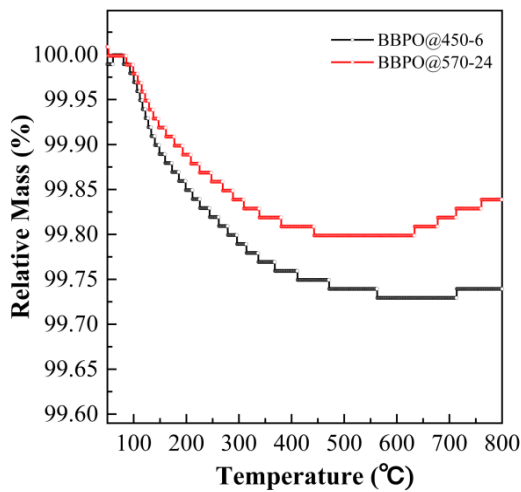

**Fig. 5** TG curves of the typical reduced samples taken under flowing O$_2$. Note that the resolution of the weight loss for this measurement is ca. 0.01wt%.

Next, we used Raman spectroscopy to examine structural changes of the thermally-treated Bi:BaBPO$_5$ with respect to the precursor used (Fig. 6). Both the precursor and treated products demonstrate a series of scattering peaks in the range of 100-1600 cm$^{-1}$, which can be assigned to the internal modes of [PO$_4$]$^{3-}$ and [BO$_4$]$^{5-}$ units and the external modes.[47] The unmodified Bi:BaBPO$_5$ demonstrates almost identical spectrum relative to the undoped BaBPO$_5$, implying that Bi incorporation does not greatly alter the host structure. Interestingly, the thermal treatment of Bi:BaBPO$_5$ leads to the occurrence of three distinct differences. First, it is found that almost all Raman peaks gradually shift to lower wavenumbers with the increased treatment duration or temperature (Fig. S2). It has revealed that the vibrational modes of P-O units in inorganic phosphates are correlated with the bond length of P-O; the increased bond length commonly leads to the shift of vibrational bands to lower wavenumbers (i.e., the birth of weaker vibrational modes).[48,49] Therefore, we conclude that the extraction of oxygen in the BaBPO$_5$ results in the increased bond lengths of B-O and P-O, probably resulting from the creation of under-coordinated B-O and P-O units. Second, the full widths at half maximum of some Raman peaks become larger with the increasing treatment harshness, in particular for the BBPO@510-24 and BBPO@570-24 samples, reflecting the increased disorder of the host because of the loss of oxygen atoms (Fig. S3). That is, the BaBPO$_5$ crystalline network is modified, although the long-range order of the crystal retains well. Third, the BBPO@510-24 and BBPO@570-24 samples display much weaker Raman scattering with respect to others, further consolidating that the topotactic reaction of Bi:BaBPO$_5$ with Al powders could disrupt the crystalline network.

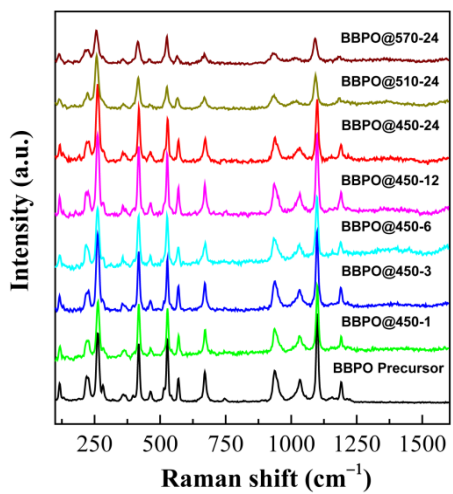

**Fig. 6** Raman spectra of the precursor and reduced samples.

The combined analyses based on TG and Raman spectroscopy lead us to conclude that the mild treatment of Bi:BaBPO$_5$ by using Al as oxygen getters creates novel metastable phases containing oxygen vacancies, i.e. the crystalline structure of BaBPO$_5$ is topotactically engineered. The direct consequence of such a transformation is the birth of Bi in an under-coordinated environment. In BaBPO$_5$ crystals, it is commonly considered that Bi ions preferentially occupy the sites of Ba$^{2+}$, thus giving rise to a coordination number of 10 with oxygen.[24] Nevertheless, Seijo and coworkers convincingly revealed that only a very low concentration of doped Bi can be incorporated into the Sr$^{2+}$ sites of the SrB4O7 matrix, because increasing Bi doping cannot increase the emission intensity.[26a] In our case, we find that the increase of Bi doping level from 0.5% to 3% also cannot improve the emission intensity (Fig. S4), which is similar to the phenomenon observed in Bi-doped SrB$_4$O$_7$. As is well known, X-ray absorption spectroscopy is a powerful technique for more reliable information about the average oxidation states and the site-symmetry changes around the absorbing atoms. To unambiguously monitor the changes in the valence and coordination environment of Bi, we next took the X-ray absorption near-edge structure (XANES) and extended X-ray absorption fine structure (EXAFS) spectra from the Bi L$_{III}$ edge of the precursor and treated phases (Fig. 7). Bi metal and Bi$_2$O$_3$ were used as reference materials. Remarkably, the onset of absorption edge gradually shifts to lower energies with the increase of treatment duration and temperature, giving an indication of decreased oxidation states of Bi (Fig. 7a). This is consistent with the expectation that a lower oxidation state of Bi is present in the modified phases. However, even the heavily-reduced product, the BBPO@570-24, exhibits the onset with a higher energy relative to Bi metal, strongly evidencing the absence of metallic state in this phase as a result that Bi ions still coordinate with oxygen. Further close inspection of the XANES spectra revealed that the precursor possesses an onset of absorption with energy larger than Bi$_2$O$_3$. Indeed, if all Bi atoms occupy the Ba$^{2+}$ sites in the BaBPO$_5$ matrix, the onset of absorption edge is expected to be shown at energy lower than that of Bi$_2$O$_3$. The deviation from such an expectation, combined with the PL result (Fig. S4), suggests that, akin to Bi-doped SrB$_4$O$_7$,[26a] only a minority of Bi atoms are at the Ba$^{2+}$ sites. We thus conclude that in the as-prepared Bi:BaBPO$_5$ phosphors, only a minority of Bi ions at the Ba$^{2+}$ sites contribute to the red PL, whereas most of Bi ions are optically inactive in both visible and NIR spectral ranges. *Then, an interesting question is reasonably proposed: where are the majority of doped Bi atoms?*

To answer this, we next carried out the analysis of the EXAFS spectra of the precursor and the reduced phases. Fig. 7b displays the Fourier transforms (FTs) of the EXAFS at the Bi sites for the precursor and reduced phases. As is well recognized, the peaks in R space correspond to interatomic distances between absorbing and surrounding atoms, after correcting for the photoelectron phase shift which causes the peak position to decrease by ca. 0.5 Å relative to the actual interatomic distances.[27] The peak intensity is connected with the average number of neighbors of a given type and its mean square bond length disorder.[21] Interestingly, we find that the precursor (i.e., the as-synthesized Bi:BaBPO$_5$) shows a smaller R value relative to Bi$_2$O$_3$. This means that most of Bi atoms are in polyhedra with short Bi-O bond lengths than those in Bi$_2$O$_3$. The ionic radii of four-coordinated P$^{5+}$ and B$^{3+}$ and ten-coordinated Ba$^{2+}$ are 0.17, 0.11, and 1.52 Å respectively.[50] In contrast, the radius of six-coordinated $Bi^{5+}$ is 0.76 Å, and the sizes of $Bi^{3+}$ with different coordination numbers are relatively large but much smaller than that of $Ba^{2+}$.[50] Thus, in any case that Bi substitutes for cations of hosts, there exists a larger size mismatch of atoms. Combined with the XANES and EXAFS results described above, we conclude that a minority of Bi atoms are indeed at $Ba^{2+}$ sits, whereas most of Bi atoms should occupy the $P^{5+}$ and/or $B^{3+}$ sites. In all cases, the structural distortion occurs in Bi-doped regions, thus relaxing the influence of size mismatch of atoms on the long-range order of the crystalline structure. However, owing to the large size mismatch between Bi and $P^{5+}$ or $B^{3+}$, Bi ions at $P^{5+}/B^{3+}$ sites are optically inactive in the as-synthesized product, given the fact that only one class of active center contributing to red PL emission could be observed.

Close examination of FTs of the EXAFS at the Bi sites provides another interesting and important clue on the local environment of Bi present (Fig. 7b). It is found that the peak intensity monotonously decreases as the increased harshness of thermal treatment, and the peak position does not show a remarkable shift. This can be attributed to the decreased number of coordination oxygen and/or increased disorder at the Bi sites in the reduced products.

Considering all experimental evidences, we conclude that the topotactic oxygen deintercalation of Bi:BaBPO$_5$ seriously alters the coordination environment of Bi, in particular for those at the $P^{5+}$ and/or $B^{3+}$ sites, resulting in the occurrence of [BiO$_x$] (x<4) units that are converted from PL-inactive, distorted [BiO$_4$] polyhedra, although the long-range order of the BaBPO$_5$ crystalline phase remains well (Fig. 8). These under-coordinated [BiO$_x$] units are optically active and contribute to the observed NIR PL, which becomes NIR-PL inactive when the concentration of oxygen vacancies is too high (Fig. 2b). Therefore, although at present it is difficult to determine exact nature of the [BiO$_x$] species, we can deduce that only [BiO$_x$] units with certain x values can contribute to the NIR emission. In marked contrast, the minority of $Bi^{2+}$ ions at the $Ba^{2+}$ sites can partially survive the structural transformation (Fig. 2a), probably owing to more robust Bi-O polyhedra; this results in the well-maintained red PL even for the BBPO@510-24 and BBPO@570-24 samples with high vacancy concentrations. The simultaneous existence of visible and NIR active centers in one phase may make energy transfer between them become possible, and the stronger red PL emission of the BBPO@570-24 sample with respect to the BBPO@510-24 sample can be attributed to the suppressed energy transfer from $Bi^{2+}$ ions to [BiO$_x$] units in the BBPO@570-24 sample in which the NIR active centers were almost totally destroyed (Fig. 2).

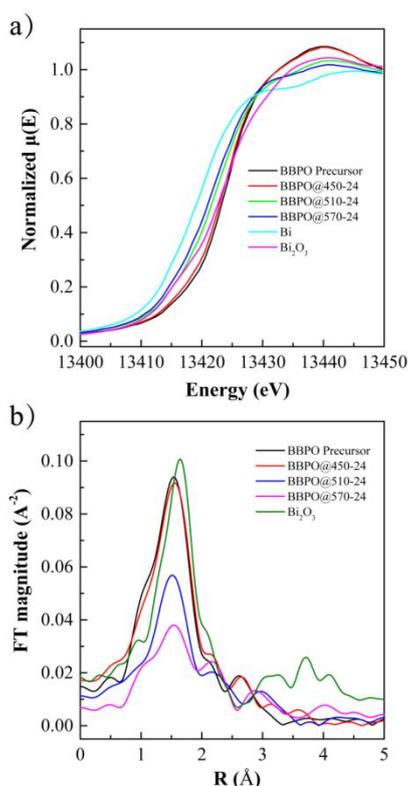

**Fig. 7** (a) Bi L$_{III}$-edge XANES spectra of the precursor, reduced samples, Bi metal and Bi$_2$O$_3$. (b) FTs of the EXAFS spectra for the precursor, reduced samples and Bi$_2$O$_3$.

If we assume that one oxygen atom is extracted from the Bi-O polyhedra, lower oxidation states of Bi with respect to that in the precursor would appear (Fig. 8). As a result, the occurrence of NIR PL can be readily associated with these defective Bi-O units bearing Bi with unconventional valences. It is noted that although a diverse array of Bi-doped NIR-luminescing crystals have been reported,[4] little attention has been paid on the investigation of coordination environments and oxidation states of Bi present.[4,21] Our results demonstrated here, coupled with an earlier report on the topotactic conversion of $Bi^{3+}$-doped crystals into NIR-luminescing materials by employing $CaH_2$ as a solid-state reducing agent,[21] indicate that the introduction of defects in Bi-doped systems is the key to achieve emerging NIR PL properties that can be commonly assigned to defective Bi-O units. Note that the PL properties of the reduced phase reported here differ from those previously reported,[21] which can be attributed to the different Bi-O units created in these systems.

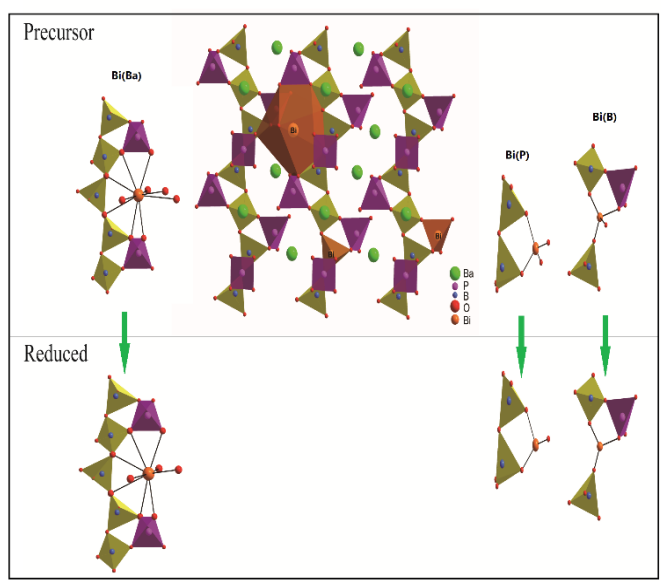

**Fig. 8** Schematic illustration of the topotactic transformation for Bi-doped $BaBPO_5$ phosphors. The upper part shows the structure of the precursor, and Bi atoms occupy the sites of $Ba^{2+}$ and $P^{5+}/B^{3+}$. The lower part displays the coordination environments of Bi at $Ba^{2+}$ and $P^{5+}/B^{3+}$ sites after removal of one oxygen atom at the apical sites in reduced products, resulting in the occurrence of under-coordinated Bi-O units. That is, a peculiar metal-oxygen-metal framework could be created by this strategy. Bi(Ba), Bi(P) and Bi(B) represent that Bi atoms occupy the Ba, P, and B sites, respectively.

## Conclusions

To summarize, we have demonstrated that Bi-doped red-emitting $BaBPO_5$ can be converted into NIR-emitting systems enabled by the topotactic removal of oxygen in the framework. Based on the combined experimental evidences gleaned from XRD, TG, Raman scattering and XAFS, we conclude that the low-temperature topochemical reduction of Bi-doped $BaBPO_5$ results in the occurrence of oxygen vacancies in the matrix, increasing the short-range disorder while preserving the long-range crystalline structure. The creation of such unique metastable networks allows access to under-coordinated Bi-O units in the matrix. XANES and EXAFS analyses help us identify that a minority of Bi atoms occupy the $Ba^{2+}$ site, and most of Bi atoms should occupy the $P^{5+}$ and/or $B^{3+}$ sites. Thus, Bi-containing, red-emitting $BaBPO_5$ should be viewed as Bi-doped $BaBPO_5$, rather than '$Bi^{2+}$'-doped ones. We also found that the topotactic treatment preferentially changes the local environment of Bi at $P^{5+}$ and/or $B^{3+}$ sites, and the created $[BiO_x]$ units, that are hard to be accessible and distinguished in materials synthesized by conventional high-temperature reactions, contribute to the NIR emission. Interestingly, such a topotactic transformation does not strongly weaken the intensity of red PL, which can be attributed to the electronic transition of $Bi^{2+}$ or molecule-like, structurally-distorted $[BiO_{10}]$ unit. This work, combined with our report on the topotactic reduction of $Bi^{3+}$-doped materials into NIR-luminescent systems by using $CaH_2$ as a solid-state reducing agent,[21] not only enrich the strategies for the controlled synthesis of novel Bi-activated

NIR-luminescent materials bearing unusual Bi-O units, but also help us gain unprecedented insight into PL mechanisms of Bi-dopd systems. The Bi-doped luminescent materials continuously surprise us, and even more than half a century of intensive research, much room still remains to advance further as well as to bring new ideas to this exciting field of science.[4] We anticipate that the low-temperature topotactic reduction strategy can be applied to the development of more novel Bi-doped luminescent materials in various forms that may find a broad range of functional applications.

## Acknowledgements

This work is financially supported by the National Natural Science Foundation of China (Grant Nos. 11574225, 51472162, 51472263, and U1532120), Jiangsu Specially Appointed Professor program (SR10900214), Natural Science Foundation of Jiangsu Province for Young Scholars (BK20140336), a project funded by the Priority Academic Program Development of Jiangsu Higher Education Institutions (PAPD), the Youth Project of National Natural Science Fund (11405256), Shanghai Technical platform for Testing and Characterization on Inorganic Materials (14DZ2292900) and Shanghai Sailing Program (16YF1413100). We appreciate the staff at the 1W2B beamline at the Beijing Synchrotron Radiation Facility for XAFS measurements. We also thank the staff at the BL14W1 beamline at the Shanghai Synchrotron Radiation Facility for XAFS measurements.